\RequirePackage{ifpdf}
\documentclass[11pt]{article}
\pdfoutput=1
\usepackage[utf8]{inputenc}
\usepackage{amsmath}
\usepackage{amssymb}
\usepackage{graphicx}
\usepackage{authblk}
\usepackage{bm}
\usepackage{cite}
\usepackage{latexsym}

\def\({\left(}
\def\){\right)}
\def\[{\left[}
\def\]{\right]}

\def\k{{\kappa}}
\def\l{{\lambda}}

\def\d{{\delta}}

\def\o{{\omega}}

\def\b{{\beta}}
\def\c{{\chi}}

\def\G{{\Gamma}}

\def\m{{\mu}}
\def\n{{\nu}}
\def\r{{\rho}}

\def\t{{\tau}}

\def\ph{{\phi}}

\def\Ph{{\Phi}}

\newcommand{\lag}{\langle}
\newcommand{\rag}{\rangle}
\newcommand{\pd}{\partial}

\newcommand{\tO}{\tilde{O}}
\newcommand{\oE}{\omega_E}
\newcommand{\no}{\nonumber \\}



  \title{\LARGE\bf D-instantons in Real Time Dynamics}

  \author[2]{Si-wen Li\thanks{siwenli@fudan.edu.cn}}
  \author[1]{Shu Lin\thanks{linshu8@mail.sysu.edu.cn}}
  \affil[1]{School of Physics and Astronomy, Sun Yat-Sen University, Zhuhai 519082, China}
  \affil[2]{Department of Physics, Center for Field Theory and Particle Physics, Fudan University, Shanghai 200433, China}

\begin{document}

\maketitle

\begin{abstract}
  Instanton is known to exist in Euclidean spacetime only. Their role in real time dynamics is usually understood as tunneling effect by Wick rotation. We illustrate other effects of instanton in holography by investigating 5d effective gravity theory of the black D3-brane-D-instanton system. The supergravity description of the D3-brane-D-instanton system is dual to the super Yang-Mills theory with topological excitations of the vacuum. We obtain Euclidean correlators in the presence of instantons by analyzing of the fluctuations of the bulk fields in the 5d effective theory. Furthermore, analytic continuation of Euclidean correlators leads to retarded correlators, which characterize real time dynamics. We find interestingly real time fluctuations of topological charge can destroy instantons and the lifetime of instanton is set by temperature. This implies instanton contribution to ``real time dynamics'' is suppressed at high temperature, which is analogous to classic field theory results that instanton contribution to ``thermodynamics'' is suppressed at high temperature.
\end{abstract}
\newpage{}


\section{Introduction}

Instantons in quantum chromodynamics (QCD) are known as topologically non-trivial excitations of the vacuum. They are known to contribution to the thermodynamics of QCD and are closely related to chiral symmetry breaking. We refer the readers to \cite{Schafer:1996wv,Gross:1980br} for comprehensive reviews. Instantons are also known to be consist of constituents called BPS monopoles or dyons. There have been continuous efforts in linking confinement and chiral symmetry breaking with instanton constituents \cite{Diakonov:2004jn,Diakonov:2007nv,Liu:2015ufa,Liu:2015jsa,Liu:2016thw,Liu:2016mrk,Liu:2016yij,Poppitz:2012sw,Poppitz:2012nz}.
While extensive literature focuses on the role of instantons in thermodynamics and phase transition, their role in real time dynamics has not received enough attention. The goal of this work is to fill the gap. Similar as instantons in quantum mechanics, instantons in QCD is generally viewed as tunneling processes. A typical question one may ask is how does the presence of instanton affects certain correlation function, e.g. a retarded correlator $G^R(\o)$, which is a genuine real time quantity. It is known that
retarded correlator is related to Euclidean counterpart $G^E(\o_E)$ by:
\begin{align}\label{GRE_intro}
G^E(\o_E)=-G^R(\o),
\end{align}
with $\o=i\o_E=2i\pi Tn$. Note that the equality holds only when $\o_E$ takes value on discrete Matsubara frequencies. In order to obtain retarded correlator with arbitrary real $\o$, analytic continuation is needed. However it is not always well defined. For example, in lattice gauge theory, $G^E$ and $G^R$ are related through spectral function. Going from $G^E$ to $G^R$ involves an ambiguity.

Holography provides us a possibility to study the effect of instantons in real time dynamics. Instantons in holographic models have been constructed as D-instanton background \cite{Liu:1999fc,Barbon:1999zp,Ghoroku:2005tf}. Effects of instantons on chiral symmetry breaking and heavy quark potential have been discussed holographically in \cite{Gwak:2012ht,Zhang:2016fdk,Zhang:2017aoc}. Thermodynamics of anisotropic instanton distribution has been studied in \cite{Mateos:2011tv,Cheng:2014qia}. While these are essentially related to Euclidean quantities, holography also allows for easy access of real time quantities. In holography, we may also work with purely imaginary $\o_E$, which can be viewed as a natural analytic continuation of \eqref{GRE_intro}. Such an approach allows for straightforward extraction of retarded correlation function from \eqref{GRE_intro}. In this paper, we illustrate the effect of instantons on retarded correlators of topological charge at finite temperature, which are measures of its fluctuations. Since instanton themselves carry topological charge, we find interestingly the fluctuations can destroy instantons in real time. The lifetime of instanton is found to be set by the temperature. Simply speaking, contribution of instanton to ``real time dynamics'' is suppressed at high temperature. This is in line with the classic field theory results: contribution of instanton to ``thermodynamics'' is suppressed at high temperature \cite{Pisarski:1980md,Shuryak:1994ay}.

The paper is organized as follows: in Section II, we review the D3-brane-D-instanton background at finite temperature. In Section III, we obtain the 5D effective action and study 5D fluctuations in D-instanton background. We proceed to calculation of two point correlators of stress tensor components, gluon condensate and topological charge in Section IV. Finally, we discuss effect of instanton in real time dynamics and conclude in Section V. 

\section{Review of D3-brane-D-instanton system}

In this section, let us briefly review the background of the black
D3-branes with D-instantons and its dual field theory.
The original example for D3-branes with D-instantons was proposed
in \cite{Liu:1999fc}, which is a deformed D3-brane solution of type
IIB supergravity by a non-trivial scalar field. In order to preserve
1/2 of supersymmetry, a Ramond-Ramond (RR) scalar charge is switched
on and balanced by the dilaton charge in this system. The resulting
solution represents a marginal \textquoteleft bound state\textquoteright
of D3-branes with smeared D-instantons i.e. D(-1)-branes. In this
paper, we focus on a finite-temperature extension of the D3-D(-1)
background in Euclidean signature \cite{Ghoroku:2005tf}. In this
background there is a RR four-form $C_{4}$ and zero-form
$\mathcal{C}$ field which couples to D3 and D(-1)-branes respectively.
In Einstein frame, the 10-dimensional (10D) supergravity action is
given as \cite{Gibbons:1995vg,Kehagias:1999iy},
\begin{equation}
S_{10\mathrm{D}}=\frac{1}{2\kappa_{10}^{2}}\int d^{10}x\sqrt{g}\left[\mathcal{R}_{10}-\frac{1}{2}\left(\partial\Phi\right)^{2}+\frac{1}{2}e^{2\Phi}\left(\partial\chi\right)^{2}-\frac{1}{2}\left|F_{5}\right|^{2}\right],\label{eq:1}
\end{equation}
where $\Phi$ is the dilaton filed, $F_{5}=dC_{4}$ is the field strength
of $C_{4}$. The original RR zero-form $\mathcal{C}$ is
defined as $\chi=i\mathcal{C}$ where $\chi$ is usually named as
axion. By setting $\chi=-e^{-\Phi}+\chi_{0}$ where $\chi_{0}$ is
a constant, the dilaton term cancels the axion term in (\ref{eq:1})
so that the dynamics in the action involves the metric and the RR
4-form $C_{4}$ only. The solution in Euclidean signature reads
\cite{Kehagias:1999iy,Ghoroku:2005tf},
\begin{align}
ds_{10}^{2}= & \frac{r^{2}}{R^{2}}\left[f\left(r\right)d\tau^{2}+\delta_{ij}dx^{i}dx^{j}\right]+\frac{1}{f\left(r\right)}\frac{R^{2}}{r^{2}}dr^{2}+R^{2}d\Omega_{5}^{2},\nonumber \\
f\left(r\right)= & 1-\frac{r_{T}^{4}}{r^{4}},\ \ \chi=-e^{-\Phi}+\chi_{0},\ \ e^{\Phi}=1+\frac{q}{r_{T}^{4}}\log\left[\frac{1}{f\left(r\right)}\right],\nonumber \\
F_{5}= & \sqrt{-2\Lambda}\left(\epsilon_{5}+\star\epsilon_{5}\right).\label{eq:2}
\end{align}
Here $R^{4}=4\pi g_{s}N_{c}l_{s}^{4}$, $\left|F_{5}\right|^{2}=\frac{1}{5!}F_{MNKPQ}F^{MNKPQ}$,
$\tau$ is the Euclidean time defined as $\tau=it$ and $\epsilon_{5}$
is the 5-form volume element and ``$\star$'' represents the Hodge
dual. The constant $q$ denotes the number density of D(-1)- branes i.e. D-instantons.
While the dilaton becomes divergent at the black hole horizon, it
does not give any effects about the thermodynamics since the bulk
observables should be computed in the Einstein frame where the on-shell
action is the same as the case of the purely black D3-brane. 

According to the AdS/CFT dictionary, the above supergravity solution
holographically describes the Euclidean $\mathcal{N}=4$ super Yang-Mills
theory with $SU\left(N\right)$ gauge symmetry in a certain non-vacuum
state. Hence the constant $q$ represents the vacuum expectation
value (VEV) in Euclidean spacetime:
\begin{align}
&O_E= \left\langle \mathrm{Tr}\mathcal{F}^2\right\rangle, \tO_E= \left\langle \mathrm{Tr}\mathcal{F}\wedge\mathcal{F}\right\rangle, \no
&O_E=-\tO_E \propto q\neq0.
\end{align}
Here we define gluon condensate $O_E$ and topological charge density $\tO_E$ in terms of the Euclidean gauge field strength $\mathcal{F}$. We use subscript $E$ to indicate that Euclidean signature is to be used. Note that sign corresponds to that of anti-instanton. Below we will loosely refer to this as instanton, which do not affect the conclusion of this paper.

The temperature of the dual field theory is given by $T=\frac{r_{T}}{\pi R^{2}}$. The non-trivial profiles of $\Phi$ and $\c$ corresponds to an effective $\theta$ term. To see this, we consider the action of a probe D3-branes with smeared D(-1)-branes,
\begin{equation}
S_{D3}=S_{DBI}+\mu_{3}\int C_{4}+\frac{i}{2}\mu_{3}\left(2\pi\alpha^{\prime}\right)^{2}\int\chi\mathcal{F}\wedge\mathcal{F}+...,\label{eq:5}
\end{equation}
where $\mu_{3}=\frac{1}{\left(2\pi\right)^{3}l_{s}^{4}}$ and $\alpha^{\prime}=l_{s}^{2}$.
The super Yang-Mills action in the dual theory comes from the leading
order expansion of the Dirac-Born-Infield (DBI) action in (\ref{eq:5}).
Since $\chi$ only depends on $r$ according to (\ref{eq:2}), the
last term in (\ref{eq:5}) give rises to a $\theta$ term
\begin{equation}
\chi\left(r\right)\sim\theta\left(r\right),\ \ \ \ \int_{\mathbb{R}^{4}}\chi\mathcal{F}\wedge\mathcal{F}=\chi\int_{\mathbb{R}^{4}}\mathcal{F}\wedge\mathcal{F}\sim\theta\int_{\mathbb{R}^{4}}\mathcal{F}\wedge\mathcal{F}.
\end{equation}
This allows us to dial instanton density $q$ as an independent parameter.

%
%
%

\section{The 5-dimensional effective theory }

\subsection{Dimensional reduction}

In this section we are going to obtain a 5-dimensional (5D) effective
gravity theory of the D3-D(-1) system and explore the 5D fluctuations
of the bulk fields. Let us start with the 10D equations of motions
for the bulk fields, which can be derived by varying the action (\ref{eq:1})\footnote{Here we have used script $\mathcal{R}$ to denote the curvature in
order to distinguish from the radius $R$ of the bulk.},
\begin{align}
\mathcal{R}_{MN}= & \frac{1}{2}\partial_{M}\Phi\partial_{N}\Phi-\frac{1}{2}e^{2\Phi}\partial_{M}\chi\partial_{N}\chi+\frac{1}{6}F_{MKPQL}F_{N}^{\ KPQL},\nonumber \\
\nabla^{2}\Phi= & -e^{2\Phi}\left(\partial\chi\right)^{2},\ \ \ \ \ \ g^{MN}\nabla_{M}\left(e^{2\Phi}\partial_{N}\chi\right)=0,\label{eq:7}
\end{align}
where the index $M,N,P,Q,L$ runs from 0 - 9. Inserting the solution
(\ref{eq:2}) into (\ref{eq:7}) and we assume all the functions only
depend on the 5D coordinates $x^{a}=\left\{ x^{\mu},r\right\} $ where
$\mu=0,1,2,3$. Hence the (\ref{eq:7}) can be rewritten as, 
\begin{align}
\mathcal{R}_{ab}^{\left(5d\right)}= & \Lambda g_{ab}+\frac{1}{2}\partial_{a}\Phi\partial_{b}\Phi-\frac{1}{2}e^{2\Phi}\partial_{a}\chi\partial_{b}\chi,\nonumber \\
0= & \frac{1}{\sqrt{g_{(5d)}}}\partial_{a}\left[\sqrt{g_{(5d)}}g^{ab}\partial_{b}\Phi\right]+e^{2\Phi}g^{ab}\partial_{a}\chi\partial_{b}\chi,\nonumber \\
0= & \frac{1}{\sqrt{g_{(5d)}}}\partial_{a}\left[\sqrt{g_{(5d)}}g^{ab}e^{2\Phi}\partial_{b}\chi\right],\nonumber \\
\mathcal{R}_{mn}= & -\Lambda g_{mn},\label{eq:8}
\end{align}
where $m,n$ runs from 6 - 9 and $\Lambda=-\frac{6}{R^{2}}$ is the
cosmological constant. The last equation is automatically satisfied
on a five-sphere, so the 5D effective action could be taken as,
\begin{equation}
S_{5D}=\frac{1}{2\kappa_{5}^{2}}\int d^{5}x\sqrt{g_{(5)}}\left[\mathcal{R}^{(5)}-\frac{1}{2}\left(\partial\Phi\right)^{2}+\frac{1}{2}e^{2\Phi}\left(\partial\chi\right)^{2}-2\Lambda\right],\label{eq:9}
\end{equation}
The reduction from 10D action to 5D effective action essentially fixes $F_5$ and metric components on $S^5$. This is justified on the gravity side as we have shown already the ansatz automatically satisfies field equation. We only turn on 5D metric components, dilaton and axion, all of which have trivial dependence on $S^5$ coordinates. This is sufficient for our purpose, since our goal is to calculate correlators among stress tensor components, $O_E$ and $\tO_E$ on the field theory side.

\subsection{The fluctuations}

Let us consider the following fluctuations in 5D effective action (\ref{eq:9}).
%
\begin{align}
g_{00} & \rightarrow g_{00}+e^{-i\omega_{E}\tau}r^{2}h_{00}\left(r\right),\nonumber \\
g_{ii} & \rightarrow g_{ii}+e^{-i\omega_{E}\tau}r^{2}h_{ii}\left(r\right),\nonumber \\
\Phi & \rightarrow\Phi+e^{-i\omega_{E}\tau}\delta\Phi\left(r\right),\nonumber \\
\chi & \rightarrow\chi+e^{-i\omega_{E}\tau}\delta\chi\left(r\right).\label{eq:10}
\end{align}
The fluctuations are taken to be specific Fourier component with Euclidean frequency $\o_E$ and are homogeneous in $\vec{x}$, respecting rotational symmetry. The rotational symmetry also allows us to set $h_{xx}=h_{yy}=h_{zz}=H$.
We have imposed the radial gauge, in which $h_{rr}=h_{\mu r}=0$. 
The equations of motion for the Euclidean fluctuations can be derived as,
\begin{align}
0= & -16r^{2}fh_{00}+16r^{2}f^{2}h_{00}-6\omega_{E}^{2}fH+2r^{3}ff^{\prime}h_{00}+r^{4}f^{\prime2}h_{00}\nonumber \\
 & +12r^{3}f^{2}h_{00}^{\prime}-r^{4}ff^{\prime}h_{00}^{\prime}+6r^{3}f^{3}H^{\prime}+3r^{4}f^{2}f^{\prime}H^{\prime}+2r^{4}f^{2}h_{00}^{\prime\prime},\nonumber \\
0= & -\omega_{E}^{2}H-8r^{2}fH+8r^{2}f^{2}H-r^{3}f^{\prime}h_{00}+2r^{3}ff^{\prime}H+r^{3}fh_{00}^{\prime}\nonumber \\
 & +8r^{3}f^{2}H^{\prime}+r^{4}ff^{\prime}H^{\prime}+r^{4}f^{2}H^{\prime\prime},\nonumber \\
0= & \frac{3\left(-f^{\prime}H+2fH^{\prime}\right)}{4f}+\frac{1}{2}\delta\Phi\Phi^{\prime}-\frac{1}{2}e^{\Phi}\Phi^{\prime}\delta\chi,\nonumber \\
0= & -2\delta\Phi\left(\omega_{E}^{2}-2r^{4}f^{2}\Phi^{\prime2}\right)-r^{4}h_{00}f^{\prime}\Phi^{\prime}+2r^{4}ff^{\prime}\delta\Phi^{\prime}+r^{4}fh_{00}^{\prime}\Phi^{\prime}\nonumber \\
 & +10r^{3}f^{2}\delta\Phi^{\prime}+3r^{4}f^{2}H^{\prime}\Phi^{\prime}+4r^{4}f^{2}e^{\Phi}\delta\chi^{\prime}\Phi^{\prime}+2r^{4}f^{2}\delta\Phi^{\prime}.\label{eq:11}
\end{align}
We only keep four equations in (\ref{eq:11}). The remaining equations are not independent since they should be satisfied automatically by the solution to the equations in (\ref{eq:11}). 
Among equations in \eqref{eq:11}, the first two are dynamical equations. The third and fourth equations are constraint equations, which can be combined to eliminate $\d\c$, giving rise to a third dynamical equation.
As a result, we need to solve only three equations for $h_{00},H,\delta\Phi$
respectively by integrating the equations from the horizon. Since
the background is Euclidean, we impose regular boundary condition.
Let us consider $\omega_{E}>0$ for now. The solutions take the following series
solutions near the horizon,
\begin{align}
h_{00}\left(r\right)= & B_{0}\left(r-1\right)^{\omega_{E}/4}\left[\frac{-24+12\omega_{E}}{2+\omega_{E}}\left(r-1\right)+\mathcal{O}\left(r-1\right)^{2}\right],\nonumber \\
H\left(r\right)= & B_{0}\left(r-1\right)^{\omega_{E}/4}\left[1-\frac{20\omega_{E}+32\omega_{E}^{2}+\omega_{E}^{3}}{8\left(2+\omega_{E}\right)^{2}}\left(r-1\right)+\mathcal{O}\left(r-1\right)^{2}\right],\nonumber \\
\delta\Phi\left(r\right)= & \left(r-1\right)^{\omega_{E}/4}\bigg[\frac{C_{0}}{-1+q\ln\left[4\left(r-1\right)\right]}-\frac{6B_{0}}{\omega_{E}\left(2+\omega_{E}\right)}\nonumber \\
 & -\frac{3\left(\omega_{E}-2\right)B_0}{8q}\left(-1+q\ln\left[4\left(r-1\right)\right]\right)+\mathcal{O}\left(r-1\right)^{1}\bigg].\label{eq:12}
\end{align}
We have set $r_{T}=1$ in the above solution for simplicity. This sets scale by having $\pi T=1$. The series solution of $h_{00}$ and
$H$ are obtained by solving the first two dynamical equations in
(\ref{eq:11}) and we find that there is only one independent solution
with normalization constant $B_{0}$. When $\Phi=0$, this is the Euclidean counterpart of the infalling solution discussed by Policastro, Son and Starinets (PSS) \cite{Policastro:2002tn} in the limit of vanishing spatial momentum.
When we have nontrivial $\Phi$, $\delta\Phi$
satisfies an inhomogeneous dynamical equation whose general solution
is the sum of homogeneous solution (proportional to $B_{0}$) and
special solution (proportional to $C_{0}$). Note that the near horizon
solution is modified from simple series solution due to the non-trivial
profile of $\Phi$. Here we organize the solution as power series in $(r-1)$, treating $\ln(r-1)$ as $\mathcal{O}((r-1)^0)$. \eqref{eq:12} is essentially all order in $\ln(r-1)$.
Integrating these solutions to the boundary, we
could obtain two independent solutions with regularity boundary
condition. The number of solutions does not match the number of independent
sources. The remaining solutions are pure gauge ones, which do not satisfy regulairty boundary condition \cite{Policastro:2002tn}. In our case, the pure gauge solution is given by,
\begin{align}
h_{\mu\nu}= & \frac{1}{r^{2}}\left(\nabla_{\mu}\xi_{\nu}+\nabla_{\nu}\xi_{\mu}\right),\nonumber \\
\delta\Phi= & \xi^{\mu}\partial_{\mu}\Phi,\nonumber \\
\delta\chi= & \xi^{\mu}\partial_{\mu}\chi.
\end{align}
The gauge function $\xi$ has to be chosen such that fluctuation fields
remain in radial gauge. The explicit expressions of the gauge function
are found to be,
\begin{align}
\xi_{\tau}= & e^{-i\omega_{E}\tau}\frac{2\left(r^{4}-1\right)D_{2}-i\omega_{E}D_{1}\sqrt{r^{4}-1}}{2r^{2}},\nonumber \\
\xi_{r}= & e^{-i\omega_{E}\tau}\frac{r}{\sqrt{r^{4}-1}}D_{1}.\label{eq:14}
\end{align}
The corresponding pure gauge solution is given by
\begin{align}
h_{00}&=f^{1/2}\(2+\frac{2}{r^4}\)D_1-\frac{\oE^2f^{1/2}}{r^2}D_1-2i\oE f D_2, \no
H&=2f^{1/2}D_1, \no
\d\Phi&=\frac{-4q D_1}{r^4f^{1/2}\(1-q\ln f\)}, \no
\d\c&=\frac{-4q D_1}{r^4f^{1/2}\(1-q\ln f\)^2}.
\end{align}
Notice that there are two independent normalization constants $D_{1,2}$
in (\ref{eq:14}) which precisely generate the remaining two solutions.
Hence we totally have four solutions at hand, they are two numerical
ones and two analytic pure gauge solutions. To be specific, let us
label the solutions by $i=I,II,III,IV$ with
\begin{align}
I: & B_{0}=1,C_{0}=0,D_{1}=D_{2}=0,\nonumber \\
II: & B_{0}=0,C_{0}=1,D_{1}=D_{2}=0,\nonumber \\
III: & B_{0}=0,C_{0}=0,D_{1}=1,D_{2}=0,\nonumber \\
IV: & B_{0}=0,C_{0}=0,D_{1}=0,D_{2}=1.\label{eq:15}
\end{align}
Then we can analyze the equations of motion near the boundary and
calculate the correlation functions among the dual operators from
the asymptotic series of the fields $h_{00}$, $H$, $\delta\Phi$ and $\delta\chi$.
We can obtain the following asymptotic behavior,
\begin{align}
h_{00}= & a_{0}+\frac{a_{1}}{r^{2}}+\frac{a_{2}}{r^{4}}+\frac{a_{3}}{r^{6}}+\frac{a_{4}}{r^{8}}+\frac{a_{h}}{r^{8}}\ln r+...,\nonumber \\
H= & b_{0}+\frac{b_{1}}{r^{2}}+\frac{b_{2}}{r^{4}}+\frac{b_{3}}{r^{6}}+\frac{b_{4}}{r^{8}}+\frac{b_{h}}{r^{8}}\ln r+...,\nonumber \\
\delta\Phi= & f_{0}+\frac{f_{1}}{r^{2}}+\frac{f_{2}}{r^{4}}+\frac{f_{h}}{r^{4}}\ln r+...,\nonumber \\
\delta\chi= & c_{0}+\frac{c_{1}}{r^{2}}+\frac{c_{2}}{r^{4}}+\frac{c_{h}}{r^{4}}\ln r+...,\label{eq:16}
\end{align}
with the recursion relations among coefficients
\begin{align}\label{recursion}
 & a_{1}=-\frac{b_{0}\omega_{E}^{2}}{2},\ a_{2}=-q\left(c_{0}-f_{0}\right)+\frac{3}{2}b_{0}-a_{0},\ a_{3}=\frac{\omega_{E}^{2}}{4}b_{0},\ a_{h}=\frac{\omega_{E}^{4}}{64}\left(b_{0}+2b_{2}\right),\nonumber \\
 & b_{1}=0,\ b_{2}=\frac{q}{3}\left(c_{0}-f_{0}\right)-\frac{1}{2}b_{0},\ b_{3}=-\frac{\omega_{E}^{2}}{12}\left(b_{0}+2b_{2}\right),\ b_{h}=\frac{\omega_{E}^{4}}{64}\left(b_{0}+2b_{2}\right),\nonumber \\
 & f_{1}=-\frac{f_{0}\omega_{E}^{2}}{4},\ f_{h}=\frac{f_{0}\omega_{E}^{4}}{16},\ c_{1}=-\frac{c_{0}\omega_{E}^{2}}{4},\ c_{h}=\frac{c_{0}\omega_{E}^{4}}{16},
\end{align}
Following holographic dictionary, we take $a_0$, $b_0$, $c_0$ and $f_0$ as sources to operators $T^{00}_E$, $T^{ii}_E$, $\tO_E$ and $O_E$ respectively. The coefficients $a_2$, $b_2$, $c_2$ and $f_2$ are corresponding VEVs. It may seems odd that $a_2$ and $b_2$ are completely determined by recursion relations. We show in Section \ref{sec_wi} that correlators among $T^{00}$ and $T^{ii}$ are completely fixed by Ward identities in the limit of vanishing spatial momentum, making $a_2$ and $b_2$ non-dynamical.
The coefficients $c_{2},f_{2}$ can not be determined
by analyzing the boundary behavior only. They are related through
the constraint equation,
\begin{align}\label{constraint}
c_{2} & =\frac{-6b_{2}+24b_{4}-3b_{h}+4qf_{2}-4q^{2}f_{0}}{4q}.
\end{align}
The coefficients $a_4$ and $b_4$ are not determined upto the order we work.
In order to calculate the correlation function, we turn on the sources for the operators and measure the corresponding VEVs. The ratio of the VEV to the sources can be defined by the following response matrix,
\begin{equation}\label{Gmat}
\left(\begin{array}{cccc}
G_{aa} & G_{ab} & G_{ac} & G_{af}\\
G_{ba} & G_{bb} & G_{bc} & G_{bf}\\
G_{ca} & G_{cb} & G_{cc} & G_{cf}\\
G_{fa} & G_{fb} & G_{fc} & G_{ff}
\end{array}\right)=\left(\begin{array}{cccc}
\frac{\pd a_{2}}{\pd a_{0}} & \frac{\pd a_{2}}{\pd b_{0}} & \frac{\pd a_{2}}{\pd c_{0}} & \frac{\pd a_{2}}{\pd f_{0}}\\
\frac{\pd b_{2}}{\pd a_{0}} & \frac{\pd b_{2}}{\pd b_{0}} & \frac{\pd b_{2}}{\pd c_{0}} & \frac{\pd b_{2}}{\pd f_{0}}\\
\frac{\pd c_{2}}{\pd a_{0}} & \frac{\pd c_{2}}{\pd b_{0}} & \frac{\pd c_{2}}{\pd c_{0}} & \frac{\pd c_{2}}{\pd f_{0}}\\
\frac{\pd f_{2}}{\pd a_{0}} & \frac{\pd f_{2}}{\pd b_{0}} & \frac{\pd f_{2}}{\pd c_{0}} & \frac{\pd f_{2}}{\pd f_{0}}
\end{array}\right).
\end{equation}
Due to operator mixing in the renormalization group (RG) flow, the off-diagonal matrix elements in the response matrix is nonvanishing. 
For any given basis solution, the response matrix satisfies
\begin{align}
\left(\begin{array}{cccc}
a_{2}^{i}\\
b_{2}^{i}\\
c_{2}^{i}\\
f_{2}^{i}
\end{array}\right)=
\left(\begin{array}{cccc}
G_{aa} & G_{ab} & G_{ac} & G_{af}\\
G_{ba} & G_{bb} & G_{bc} & G_{bf}\\
G_{ca} & G_{cb} & G_{cc} & G_{cf}\\
G_{fa} & G_{fb} & G_{fc} & G_{ff}
\end{array}\right)
\left(\begin{array}{cccc}
a_{0}^{i}\\
b_{0}^{i}\\
c_{0}^{i}\\
f_{0}^{i}
\end{array}\right),
\end{align}
for $i=I,\,II,\,III,\,IV$. Since we have in total four basis solutions, we can use them to calculate the response matrix efficiently as,
\begin{equation}\label{Gab_def}
\left(\begin{array}{cccc}
G_{aa} & G_{ab} & G_{ac} & G_{af}\\
G_{ba} & G_{bb} & G_{bc} & G_{bf}\\
G_{ca} & G_{cb} & G_{cc} & G_{cf}\\
G_{fa} & G_{fb} & G_{fc} & G_{ff}
\end{array}\right)=\left(\begin{array}{cccc}
a_{2}^{I} & a_{2}^{II} & a_{2}^{III} & a_{2}^{IV}\\
b_{2}^{I} & b_{2}^{II} & b_{2}^{III} & b_{2}^{IV}\\
c_{2}^{I} & c_{2}^{II} & c_{2}^{III} & c_{2}^{IV}\\
f_{2}^{I} & f_{2}^{II} & f_{2}^{III} & f_{2}^{IV}
\end{array}\right)\left(\begin{array}{cccc}
a_{0}^{I} & a_{0}^{II} & a_{0}^{III} & a_{0}^{IV}\\
b_{0}^{I} & b_{0}^{II} & b_{0}^{III} & b_{0}^{IV}\\
c_{0}^{I} & c_{0}^{II} & c_{0}^{III} & c_{0}^{IV}\\
f_{0}^{I} & f_{0}^{II} & f_{0}^{III} & f_{0}^{IV}
\end{array}\right)^{-1}.
\end{equation}
Most of the matrix elements on the RHS of \eqref{Gab_def} are known analytically. From solutions $II$, $III$ and $IV$, we easily obtain
\begin{align}
&a_0^{II}=0,\quad b_0^{II}=0,\quad a_2^{II}=0,\quad b_2^{II}=0, \quad b_4^{II}=0, \no
&a_0^{III}=2,\quad b_0^{III}=2, \quad f_0^{III}=0,\quad a_2^{III}=1, \quad b_2^{III}=-1,\quad f_2^{III}=-4q, \quad b_4^{III}=-\frac{1}{4}, \no
&a_0^{IV}=-2i\o,\quad b_0^{IV}=0, \quad f_0^{IV}=0,\quad a_2^{IV}=2i\o, \quad b_2^{IV}=0,\quad f_2^{IV}=0, \quad b_4^{IV}=0.
\end{align}
With recursion relations \eqref{recursion} and constraint \eqref{constraint}, we can fix most entries of the response matrix as,
\begin{equation}
\left(\begin{array}{cccc}
G_{aa} & G_{ab} & G_{ac} & G_{af}\\
G_{ba} & G_{bb} & G_{bc} & G_{bf}\\
G_{ca} & G_{cb} & G_{cc} & G_{cf}\\
G_{fa} & G_{fb} & G_{fc} & G_{ff}
\end{array}\right)=\left(\begin{array}{cccc}
-1 & 3/2 & -q & q\\
0 & -1/2 & q/3 & -q/3\\
0 & -2q & \times & \times\\
0 & -2q & \times & \times
\end{array}\right).\label{eq:21}
\end{equation}
The entries marked with ``$\times$'' represent the value which has to be determined numerically. In fact, the four undetermined entries are not all independent. We can show that
there is a less obvious identity among them
\begin{align}\label{G_identity}
G_{cf}-G_{fc}+2q=G_{ff}-G_{cc}.
\end{align}
Finally we remark on one important property of response matrix under sign flip of $\oE$. Up to now we have considered $\oE>0$, the situation with $\oE<0$ is easy to analyze: the regularity condition requires the solutions near horizon $\sim (r-1)^{-\oE/4}$. It follows that solutions $I$ and $II$ remain unchanged under sign flip. On the other hand, the explicit pure gauge solutions show that $III$ is also unchanged, while $IV$ changes sign. Combining all these with recursion relations \eqref{recursion} and constraint \eqref{constraint}, we can show that the response matrix is also unchanged under sign flip of $\oE$. This will be useful in the following.

\section{Euclidean and Retarded Correlators}

\subsection{Euclidean correlators}

The D3-brane-D-instanton background is found in Euclidean spacetime
with non-trivial profiles of dilaton and axion which is consistent
with the field theory expectation since instanton only exists in Euclidean
field theory. The effect of instanton to real world physics is usually
understood as tunneling process. We will calculate correlators
among stress tensor, glueball and topological charge density in this
section, which are dual to the perturbation of metric, the dilaton
and axion respectively as discussed in the previous sections. Notice
that the resulting correlators are all Euclidean in our current
analysis of the bulk gravity. In the next section, we will study the
counterpart in Minkowskian spacetime.
%

We show in appendix how to obtain correlators 
among $T_{E}^{00},T_{E}^{ii},O_{E},\tilde{O}_{E}$
from (\ref{eq:21}) and (\ref{eq:26}). Here sum over $i$ is assumed in $T_E^{ii}$ as this is the quantity coupled to isotropic metric perturbation $H(r\to\infty)$. Respectively, the correlators are given by,
\begin{align}
G_{00,00}^E&=\int d\tau d^{3}xe^{i\omega_{E}\tau}\left\langle T_{E}^{00}\left(\tau,x\right)T_{E}^{00}\left(0\right)\right\rangle = 3,\nonumber\\
G_{00,ii}^E&=\int d\tau d^{3}xe^{i\omega_{E}\tau}\left\langle T_{E}^{00}\left(\tau,x\right)T_{E}^{ii}\left(0\right)\right\rangle = 3,\nonumber\\
G_{ii,jj}^E&=\int d\tau d^{3}xe^{i\omega_{E}\tau}\left\langle T_{E}^{ii}\left(\tau,x\right)T_{E}^{jj}\left(0\right)\right\rangle = -9,\nonumber\\
G_{00,O}^E&=\int d\tau d^{3}xe^{i\omega_{E}\tau}\left\langle T_{E}^{00}\left(\tau,x\right)O_{E}\left(0\right)\right\rangle =  4q,\nonumber \\
G_{00,\tO}^E&=\int d\tau d^{3}xe^{i\omega_{E}\tau}\left\langle T_{E}^{00}\left(\tau,x\right)\tilde{O}_{E}\left(0\right)\right\rangle =  -4q,\nonumber \\
G_{ii,O}^E&=\int d\tau d^{3}xe^{i\omega_{E}\tau}\left\langle T_{E}^{ii}\left(\tau,x\right)O_{E}\left(0\right)\right\rangle =  -4q,\nonumber \\
G_{ii,\tO}^E&=\int d\tau d^{3}xe^{i\omega_{E}\tau}\left\langle T_{E}^{ii}\left(\tau,x\right)\tilde{O}_{E}\left(0\right)\right\rangle =  4q,\nonumber \\
G_{OO}^E&=\int d\tau d^{3}xe^{i\omega_{E}\tau}\left\langle O_{E}\left(\tau,x\right)O_{E}\left(0\right)\right\rangle = 4G_{ff},\nonumber \\
G_{\tO\tO}^E&=\int d\tau d^{3}xe^{i\omega_{E}\tau}\left\langle \tilde{O}_{E}\left(\tau,x\right)\tilde{O}_{E}\left(0\right)\right\rangle = -4G_{cc},\nonumber \\
G_{O\tO}^E&=\int d\tau d^{3}xe^{i\omega_{E}\tau}\left\langle O_{E}\left(\tau,x\right)\tilde{O}_{E}\left(0\right)\right\rangle = -2G_{cf}+2G_{fc}-4q.\label{eq:28}
\end{align}
Note that all correlators are in unit of $P=\frac{\pi^2N^2T^4}{8}$, which is the pressure of plasma, as we have shown in the appendix.
The correlators among $T_{E}^{00}$ and $T_{E}^{ii}$ are trivial. We will see shortly that they are simply fixed by Ward identities derived by PSS \cite{Policastro:2002tn}.
The cross correlators among $T_{E}^{00},T_{E}^{ii}$ and $O_{E},\tilde{O}_{E}$ are
fixed by a new set of Ward identities. 
The correlators among $O_{E}$ and stress tensor components
arises naturally from the fact that spacetime integral of $\tilde{O}_{E}$
is a topologically protected quantity,
\begin{equation}
\delta\int d\tau d^{3}x\sqrt{g}\tilde{O}_{E}\left(x\right)=\int d\tau d^{3}x\sqrt{g}\left[\frac{1}{2}g^{\mu\nu}\delta g_{\mu\nu}\tilde{O}_{E}\left(x\right)+\delta\tilde{O}_{E}\right]=0.
\end{equation}
Notice that $\lag\tO_E(x)\rag=4q$. 
By taking $\d g_{00}=h_{00}(r\to\infty)$ and $\d g_{ii}=0$, we
can obtain the response of $\tilde{O}_{E}$ to the metric perturbation $\d g_{00}$,
\begin{align}
\frac{\d \lag\tO_E(x)\rag}{\d g_{00}(y)}=-2q \d^4(x-y).
\end{align}
What we obtain above is nothing but the correlator $\lag \tilde{O}_{E}(x)T_E^{00}(y)\rag$. It takes a familiar form in momentum space\footnote{The coupling $\int d^4x\frac{1}{2}h_{\m\n}T^{\m\n}$ brings in an additional factor of $2$.}
\begin{align}
\int d\tau d^{3}xe^{i\omega_{E}\tau}\left\langle \tilde{O}_{E}\left(\tau,x\right)T_{E}^{00}\left(0\right)\right\rangle = & -4q. 
\end{align}
Similarly procedure with $\d g_{00}=0$ and $\d g_{ii}=H(r\to\infty)$ leads to the correlator:
\begin{align}
\int d\tau d^{3}xe^{i\omega_{E}\tau}\left\langle \tilde{O}_{E}\left(\tau,x\right)T_{E}^{ii}\left(0\right)\right\rangle = & 4q.
\end{align}
It is not difficult to see that they are equivalent to the corresponding correlators in \eqref{eq:28} upon using complex conjugation. 

Furthermore, we can obtain from \eqref{G_identity} the following relation among Euclidean correlators
\begin{align}
G^E_{O\tO}+2q+\frac{1}{2}\(G^E_{OO}-G^E_{\tO\tO}\)=0.
\end{align}
%

\subsection{Ward identities}\label{sec_wi}

In this section, we will show that the apparently trivial
form of the correlators are in fact consequence of Ward identities following from diffeomorphism invariance and conformal invariance of the action:
%
\begin{subequations}
\begin{align}
&S\left[g_{\mu\nu}^{4D},\phi^{4D},\chi^{4D}\right]=S\left[g_{\mu\nu}^{4D}+\nabla_{\mu}\xi_{\nu}^{4D}+\nabla_{\nu}\xi_{\mu}^{4D},\phi^{4D}+\xi^{\mu}\partial_{\mu}\phi^{4D},\chi^{4D}+\xi^{\mu}\partial_{\mu}\chi^{4D}\right], \label{eq:31} \\
&S\left[g_{\mu\nu}^{4D},\phi^{4D},\chi^{4D}\right]=S\left[\Omega^{2}g_{\mu\nu}^{4D},\phi^{4D},\chi^{4D}\right]. \label{wi_conformal}
\end{align}
\end{subequations}
We have used the label ``$4D$'' to indicate that all the variables
in (\ref{eq:31}) are four-dimensional i.e. they are sources corresponding
to $a_{0},b_{0},c_{0},f_{0}$. Varying \eqref{eq:31} with respect to $\xi_{\mu}^{4D}\left(x\right)$ and \eqref{wi_conformal} with respect to $\Omega$, we obtain,
\begin{subequations}
\begin{align}
&-\nabla_{\mu}T_{E}^{\mu\nu}\left(x\right)+O_{E}(x)\nabla^{\nu}\phi^{4D}\left(x\right)+\tilde{O}_{E}(x)\nabla^{\nu}\chi^{4D}\left(x\right)=0, \\
&g_{\m\n}\lag T^{\m\n}\rag=0. \label{eq:32}
\end{align}
\end{subequations}
Varying \eqref{eq:32} with respect to $g_{\l\r}$ and setting $g_{\m\n}=\d_{\m\n}$, $\ph^{4D}=\c^{4D}=0$, we obtain
\begin{align}\label{wi_PSS}
&q_\m\(G_E^{\m\n,\l\r}(q)+\d^{\n\l}\lag T_E^{\m\r}\rag+\d^{\n\r}\lag T_E^{\m\l}\rag-\d^{\m\n}\lag T_E^{\l\r}\rag\). \no
&\d_{\m\n}G_E^{\m\n,\l\r}=-2\lag T_E^{\l\r}\rag.
\end{align}
The derivation is parallel to the one in \cite{Policastro:2002tn} except that our background metric is flat Euclidean. We can show that in fact \eqref{wi_PSS} alone is enough to fix correlators $G^{00,00}_E$, $G^{00,ii}_E$ and $G_E^{ii,jj}$. Note that $q_\m=(\oE,{\bf 0})$. We obtain from \eqref{wi_PSS}
\begin{align}
&\oE\(G_E^{00,00}+\lag T_E^{00}\rag\)=0, \no
&\oE\(G_E^{00,ii}-\lag T_E^{ii}\rag\)=0, \no
&G_E^{00,jj}+G_E^{ii,jj}=-2\lag T_E^{ii}\rag,
\end{align}
where we always assume summation over repeated indices. Using $\lag T^{00}_E\rag=-\lag T^{ii}_E\rag=-3P$, we obtain
\begin{align}
G_E^{00,00}=3P,\quad G_E^{00,ii}=3P, \quad G_E^{ii,jj}=-9P.
\end{align}
It indeed agrees with \eqref{eq:28} and explains the non-dynamical origin of the correlators.

Now we derive a new set of Ward identities involving $O_E$ and $\tO_E$. Vary (\ref{eq:32}) with respect to $\phi^{4D}\left(y\right)$
and $\chi^{4D}\left(y\right)$ and setting $g_{\m\n}=\d_{\m\n}$, $\ph^{4D}=\c^{4D}=0$, we obtain from \eqref{eq:31}:
\begin{align}
&-q_\m G_E^{\m\n,O}-q^\n\lag O_E\rag=0, \no
&-q_\m G_E^{\m\n,\tO}-q^\n\lag \tO_E \rag=0.
\end{align}
Note that $\left\langle O_{E}\left(x\right)\right\rangle =-\left\langle \tilde{O}_{E}\left(x\right)\right\rangle =-4q$,
we obtain immediately
\begin{align}\label{correlator_00}
G_E^{00,O} = & 4q\nonumber \\
G_E^{00,\tO} = & -4q.
\end{align}
Similarly, we can obtain from \eqref{wi_conformal}
\begin{align}\label{c_ward}
\d_{\mu\nu}(x)\lag T^{\mu\nu}_E\left(x\right)O_E\left(0\right)\rag = & 0,\nonumber \\
\d_{\mu\nu}(x)\lag T^{\mu\nu}_E\left(x\right)\tilde{O}_E\left(0\right)\rag = & 0.
\end{align}
Fourier transform of \eqref{c_ward} gives
\begin{align}
&G_E^{00,O}+G_E^{ii,O}=0 \no
&G_E^{00,\tO}+G_E^{ii,\tO}=0.
\end{align}
Combining with \eqref{correlator_00}, we obtain
\begin{align}
G_E^{ii,O} = & -4q\nonumber \\
G_E^{ii,\tO} = & 4q.
\end{align}
All the cross correlators are fixed by Ward identities in the limit of vanishing spatial momentum.

\subsection{Analytic Continuation}

To see how the presence of instanton affects real time dynamics, we need to know corresponding correlators in Minkowski spacetime. Conventional holographic approach is to directly work with Minkowski background, which automatically gives real time correlator. However this is not applicable to D-instanton backgrund: A naive Wick rotation of the background would lead to purely imaginary vev for $O$ and $\tO$. This is in fact consistent with the field theory expectation that instanton exists only in Euclidean space. So only Euclidean correlator is well defined. Nevertheless, analytic continuation between Euclidean correlator and real time correlator is possible in field theory. We will use the following identity between retarded correlator and Euclidean correlator:
\begin{align}\label{GRE}
G^R(i\o_E)=-G^E(\o_E),
\end{align}
where $\o_E$ takes values of Matsubara frequency $\o_E=2\pi Tn$. A clean derivation of \eqref{GRE} can be found for example in \cite{Moore:2010jd}. In \eqref{GRE}, $G^R$ and $G^E$ are defined with the same operator. It applies to the case of dilaton, whose explicit Euclidean and Minkowskian correlators are given by 
\begin{align}\label{GRE_def}
&G^R_{OO}(\o)=-i\int dtd^3xe^{i\o t}\lag [O(t,x), O(0)]\rag, \no
&G^E_{OO}(\o_E)=\int d\t d^3xe^{i\o_E\t}\lag O_E(\t,x) O_E(0)\rag,
\end{align}
with $O=O_E$.
Minkowskian and Euclidean times are related by $it=\t$, correspondingly $\o=i\o_E$.
We have already evaluated the Euclidean correlators in D-instanton background. Through \eqref{GRE} they give us automatically the retarded correlators, albeit defined only on $\o=i2\pi Tn$. To extend the results to real frequency, we need to analytically continue \eqref{GRE}. There is a natural continuation:
\begin{align}\label{GRE_ac}
G^R_{OO}(\o)=-G^E_{OO}(-i\o),
\end{align}
with $\o$ being real. \eqref{GRE_ac} indicates that to obtain retarded correlator with real $\o$, we need to evaluate Euclidean correlator with purely imaginary frequency $-i\o$. 
Note that we have been working with $\o_E>0$ for the evaluation of Euclidean correlators. The horizon solution fixed by regularity condition behaves as $\sim(r-1)^{\o_E/4}$. The horizon solution can be extended to the the right half plane $Re\o_E\ge0$. Using the relation $\o_E=-i\o$, it is mapped to the upper half plane of $\o$. Note that the horizon solution in terms of $\o$ is precisely infalling wave type $\sim(r-1)^{-i\o/4}$. Had we started by extending the left half plane of $\o_E$, where the horizon solution behaves as $\sim\(r-1\)^{-\o_E/4}$, we would not obtain infalling horizon solution by the same mapping $\o_E=-i\o$.
To further confirm the prescription, we show that the analytic continuation \eqref{GRE_ac} holds for known examples of Super Yang-Mills plasma with $q=0$ where direct calculation of both Euclidean and retarded correlators are possible. In this case, the EOM of dilaton decouples from metric perturbation. It is given by:
\begin{align}
  -2\o_E^2\d\Ph+r^3\(2rff'\d\Ph'+f^2\(10\d\Ph'+2r\d\Ph''\)\)=0
\end{align}
The EOM of dilaton in Minkowski background is given by the mapping $\o_E=-i\o$:
\begin{align}
  2\o^2\d\Ph+r^3\(2rff'\d\Ph'+f^2\(10\d\Ph'+2r\d\Ph''\)\)=0
\end{align}
Note that regular horizon solution in the Euclidean case is also mapped to the infalling horizon solution in the Minkowskian case. Taking into the account the convention in the definition \eqref{GRE_def}, we readily confirm the equality \eqref{GRE_ac} for $Re\o_E\ge0$.

Next, we discuss auto-correlator of $\tO$ defined as
\begin{align}\label{GRE_def_tO}
&G^R_{\tO\tO}(\o)=-i\int dtd^3xe^{i\o t}\lag [\tO(t,x), \tO(0)]\rag, \no
&G^E_{\tO\tO}(\o_E)=\int d\t d^3xe^{i\o_E\t}\lag \tO_E(\t,x) \tO_E(0)\rag.
\end{align}
In contrast to the above case, $\tO$ and $\tO_E$ differ by a factor.
To see this, we write down the field theory definitions for $\tO$ and $\tO_E$ (for comparison we also include $O$ and $O_E$):
\begin{align}
O_E=tr{\cal F}_E^2=tr{\cal E}_E^2+tr{\cal B}_E^2,\quad O=tr{\cal F}^2=-tr{\cal E}^2+tr{\cal B}^2, \no
\tO_E=tr{\cal F}_E\wedge {\cal F}_E=tr{\cal E}_E\cdot {\cal B}_E,\quad \tO=tr{\cal F}\wedge {\cal F}=tr{\cal E}\cdot {\cal B},
\end{align}
with ${\cal E}$ and ${\cal B}$ being chromo electric and magnetic fields. 
Note that $it=\t$. We obtain $\pd_t=i\pd_\t$, thus ${\cal E}_E=-i{\cal E}$ and ${\cal B}_E={\cal B}$. It is then straightforward to verify
\begin{align}
O_E=O,\quad \tO_E=i\tO.
\end{align}
Consequently, the analytic continuation for auto-correlator of $\tO$ involves an additional minus sign.
\begin{align}\label{RE_holo}
&G^R_{\tO\tO}(\o)=G_{\tO\tO}^E(-i\o).
\end{align}
We can confirm the prescription at $q=0$ by the following indirect comparison.
Note that retarded correlators for $O$ and $\tO$ are degenerate in Minkowskian D3 brane background because the dual dilaton and axion satisfy the same equation of motion \cite{Son:2002sd}. Therefore we would obtain from \eqref{RE_holo}
\begin{align}\label{chain}
G_{\tO\tO}^E(-i\o)=G^R_{\tO\tO}(\o)=G^R_{OO}(\o)=-G_{OO}^E(-i\o).
\end{align}
Below we confirm \eqref{chain}, where both sides can be calculated with \eqref{eq:28}.
Note that axion and dilaton also satisfy the same EOM in Euclidean background, which simply gives $G_{ff}=G_{cc}$. However, the kinetic terms of dilaton and axion differ in sign. The sign difference is reflected in the equations of $G^E_{OO}$ and $G^E_{\tO\tO}$ in \eqref{eq:28}. Thus we find $G_{\tO\tO}^E(-i\o)=-G_{OO}^E(-i\o)$, consistent with \eqref{RE_holo}.

Finally, we discuss the remaining correlators among stress tensor components. Note that, under Wick rotation, the Euclidean and Minkowskian operators are related by
\begin{align}\label{wr_stress}
  T^{00}_E=-T^{00},\quad T^{ii}_E=T^{ii}.
\end{align}
It follows that the analytic continuation is modified to
\begin{align}\label{ac_stress}
  &G_R^{00,00}(\o)=-G_E^{00,00}(-i\o),\no
  &G_R^{00,ii}(\o)=G_E^{00,ii}(-i\o),\no
  &G_R^{ii,jj}(\o)=-G_E^{ii,jj}(-i\o).
\end{align}
Since we already have explicit results for Euclidean correlator among stress tensor components, we can make direct comparison with their Minkowskian counterpart in PSS \cite{Policastro:2002tn}. Setting spatial momentum to zero in their case and noting that our $G^R$ corresponds to their $G$, we readily confirm the correctness of the analytic continuation \eqref{ac_stress}.

To proceed with the case $q\ne0$, we simply apply the above prescription: we numerically integrate the horizon solution \eqref{eq:12} with $\o_E=-i\o$ for $\o>0$ to the boundary and match to asymptotic series \eqref{eq:16} to determine the response matrix using \eqref{Gab_def}. With elements of response matrix, we can use \eqref{eq:28} to determine the boundary correlator.

\subsection{Results on Retarded Correlators}

Now we are ready to use \eqref{RE_holo} to study retarded correlators for $O$ and $\tO$ respectively. The retarded correlators are readily obtained from \eqref{eq:28} as
\begin{align}\label{Goo_dict}
G^R_{OO}(\o)&= -4G_{ff}(-i\o),\nonumber \\
G^R_{\tO\tO}(\o)&=-4G_{cc}(-i\o),
\end{align}
The results are known for the case $q=0$, where $G^R_{OO}$
and $G^R_{\tO\tO}$ are degenerate. Our numerical results indicate the degeneracy is still true when $q\ll \o,\;\;q\ll T$. Furthermore, in the regime $\o\ll T$, the correlators display diffusive behavior as
\begin{equation}
G^R_{OO}=G^R_{\tO\tO}=-i\omega\Gamma_{\mathrm{CS}}/T,\label{eq:38}
\end{equation}
where $\Gamma_{\mathrm{CS}}$ is the diffusion constant of the Chern-Simons
(CS) number \cite{Son:2002sd}.

\begin{figure}[t]
  \centering
\includegraphics[width=6cm]{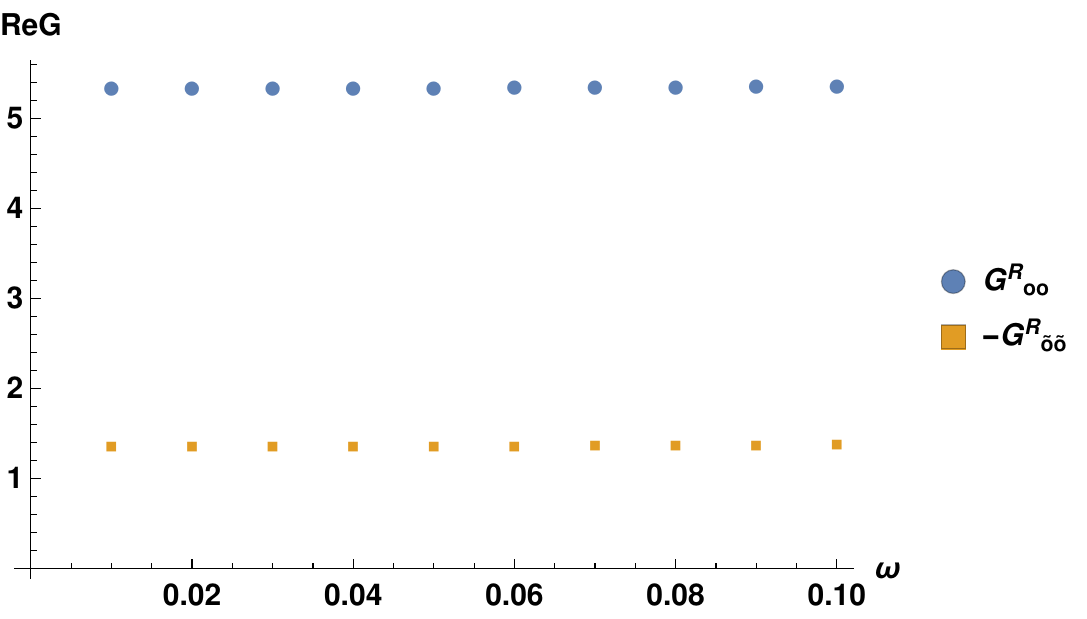}
\includegraphics[width=6cm]{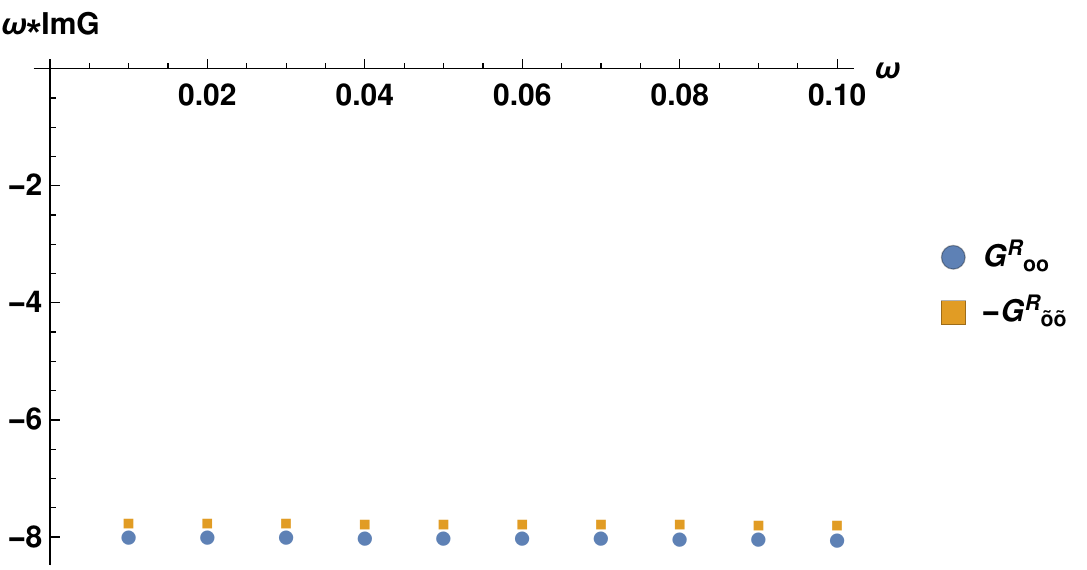}
\centering
\includegraphics[width=6cm]{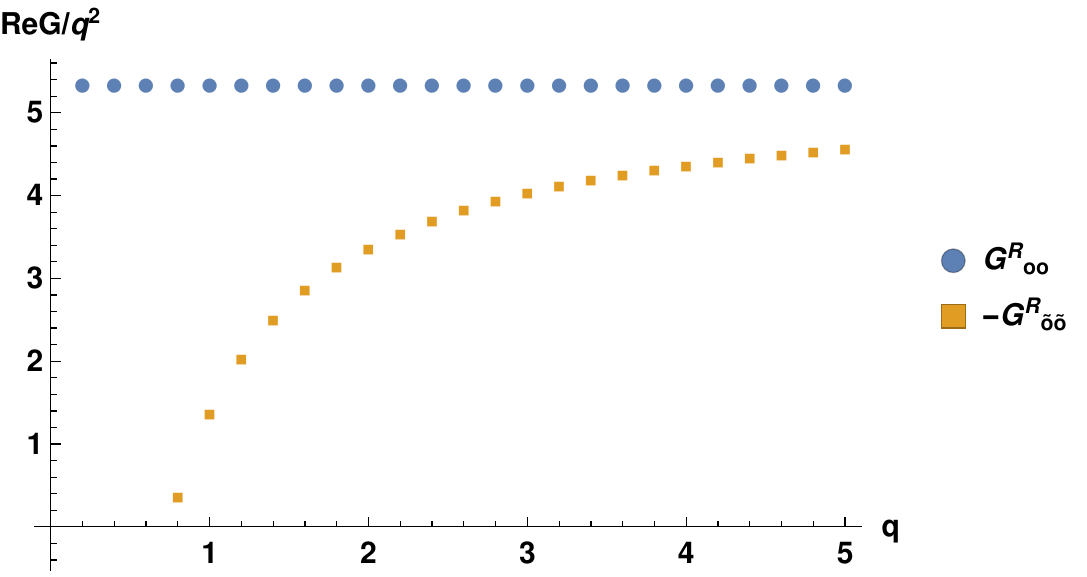}
\includegraphics[width=6cm]{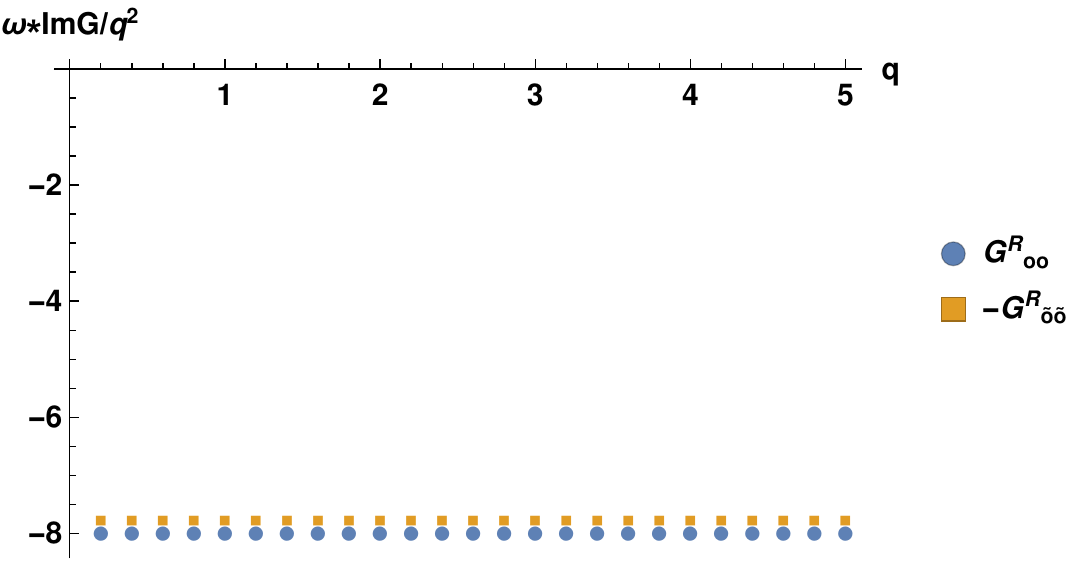}
\caption{\label{fig:2}The top panel show the real and imaginary parts of $G^R_{OO}$ and $-G^R_{\tilde{O}\tilde{O}}$ in unit of $P$ at $q/\left(\pi T\right)^4=1$ as a function of $\omega$. In low frequency regime, the real parts of $G^R_{OO}$ and $-G^R_{\tO\tO}$ differ and are almost independent of $\o$. On the other hand, the imaginary parts of $G^R_{OO}$ and $G^R_{\tilde{O}\tilde{O}}$ are almost indistinguishable. We use an offset to guide the eyes. The numerical results indicate $ImG^R_{OO}\simeq ImG^R_{\tO\tO}\sim 1/\o$. The bottom panels show the $q$ dependence of the real and imaginary parts of $G^R_{OO}$ and $-G^R_{\tO\tO}$ in unit of $P$ at $\o/\(\pi T\)=0.01$. The numerical results indicate $ReG^R_{OO}\sim q^2$ and $ImG^R_{OO}\simeq ImG^R_{\tO\tO}\sim q^2$. The $q^2$ dependence extends to rather large value of $q$.}
\end{figure}

\begin{figure}[t]
  \centering
\includegraphics[width=6cm]{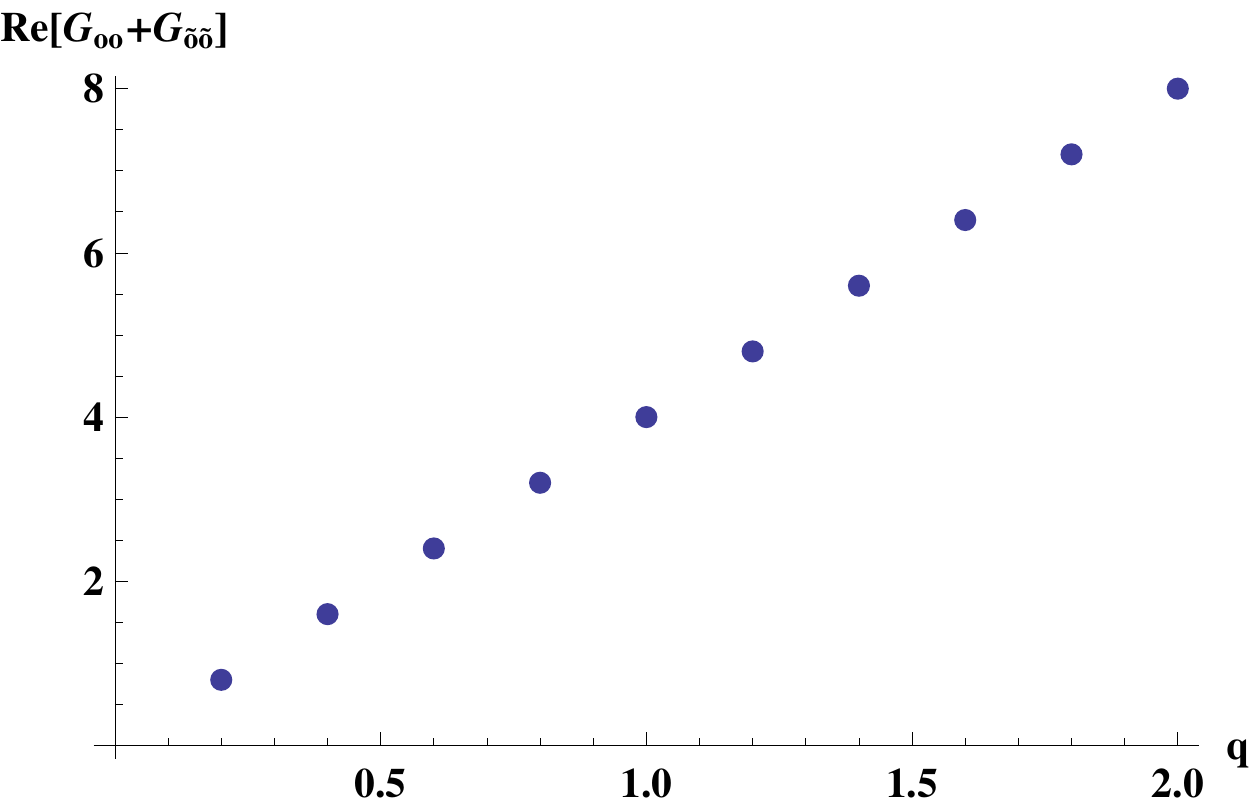}
\includegraphics[width=6cm]{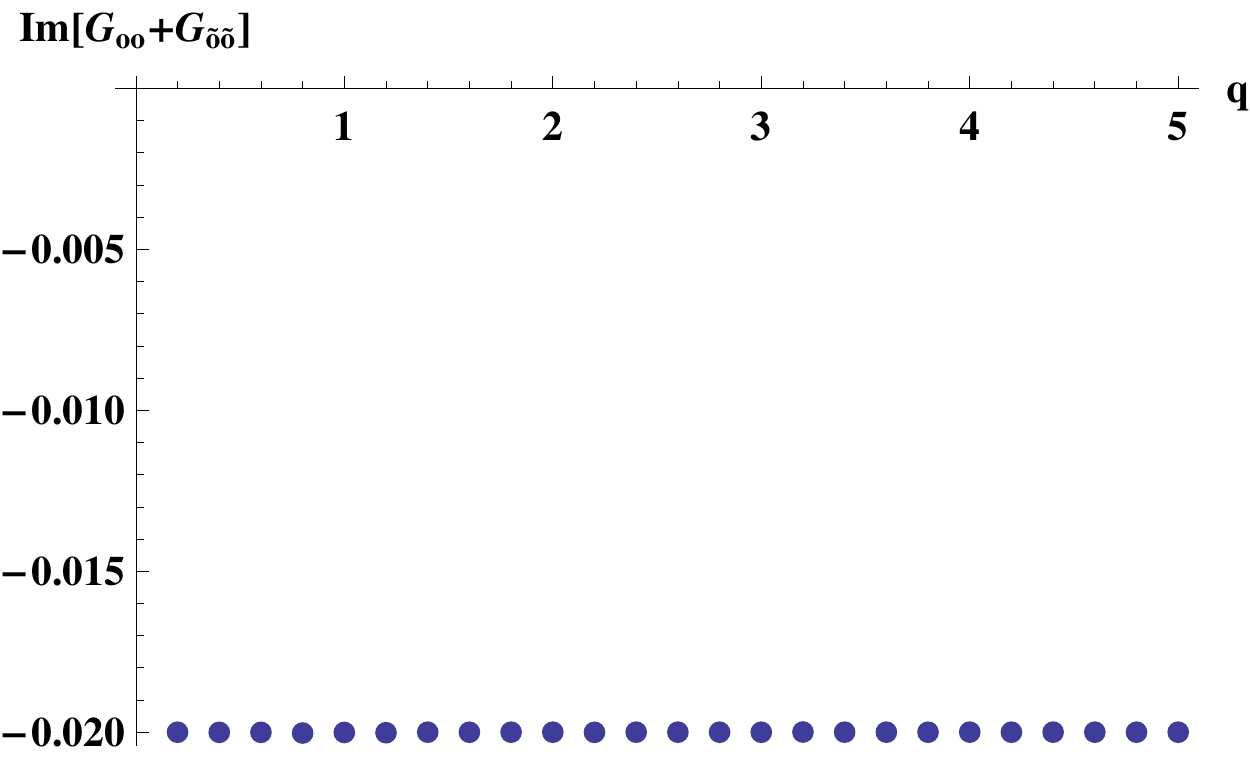}
\caption{\label{fig:3}The real and imaginary part of $G^R_{OO}+G^R_{\tO\tO}$ in unit of $P$ as a function of $q$ at $\o/\(\pi T\)=0.01$. The real part shows a clear linear dependence, while the imaginary part is almost $q$-independent.}
\end{figure}

We are interested in the regime $q\sim O(T)$. This is where instanton effect becomes significant. We first look at the regime $\o\ll T$. In Figure \ref{fig:2} we show the
$\omega$ and $q$ dependencies of the real and imaginary parts of
$G^R_{OO}$ and $-G^R_{\tO\tO}$. We find the imaginary parts of $G^R_{OO}$ and $-G^R_{\tO\tO}$ are almost indistinguishable. Note that this is in contrast to the degeneracy of $G^R_{OO}$ and $G^R_{\tO\tO}$ (without minus sign) in the regime $q\ll T$.
The numerical results also indicate the following scaling relation
\begin{align}\label{scaling}
&ReG^R_{OO}\sim q^2 ,\no
&ImG^R_{OO}\simeq -ImG^R_{\tO\tO}\sim q^2/\o.
\end{align}
We further take a closer look at the $q$-dependence of $G^R_{OO}+G^R_{\tO\tO}$ in Figure \ref{fig:3}. For the real part, Fig \ref{fig:3} shows the difference of $ReG^R_{OO}$ and $-ReG^R_{\tO\tO}$ as seen in Fig \ref{fig:2} is linear in $q$, suggesting the following scaling behavior: $ReG^R_{\tO\tO}\sim q^2+\# q$. For the imaginary part, Fig \ref{fig:3} shows there is in fact a small difference not visible in Fig \ref{fig:2}.

\begin{figure}[t]
  \centering
\includegraphics[width=6cm]{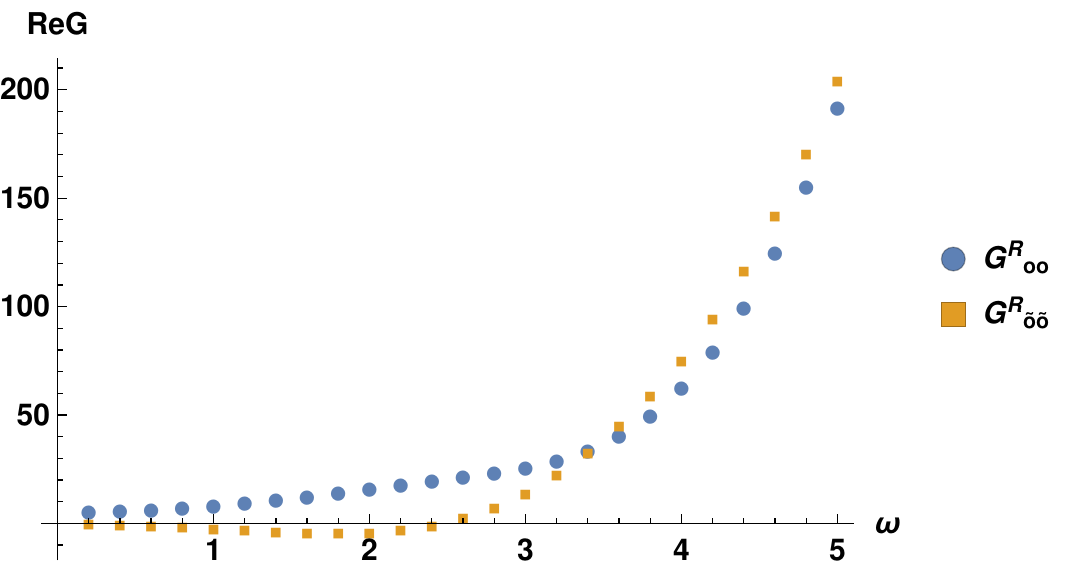}
\includegraphics[width=6cm]{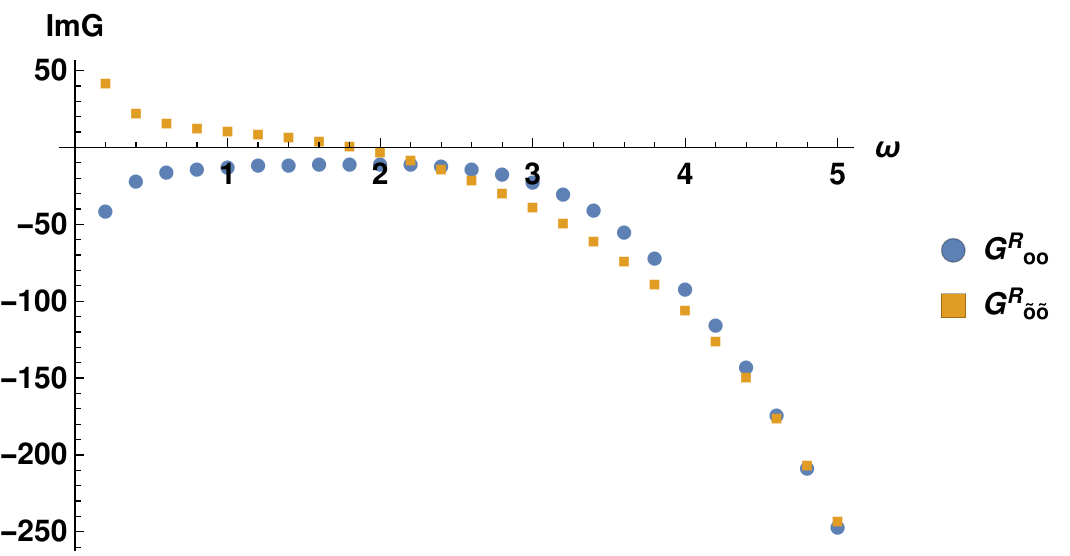}
\centering
\includegraphics[width=6cm]{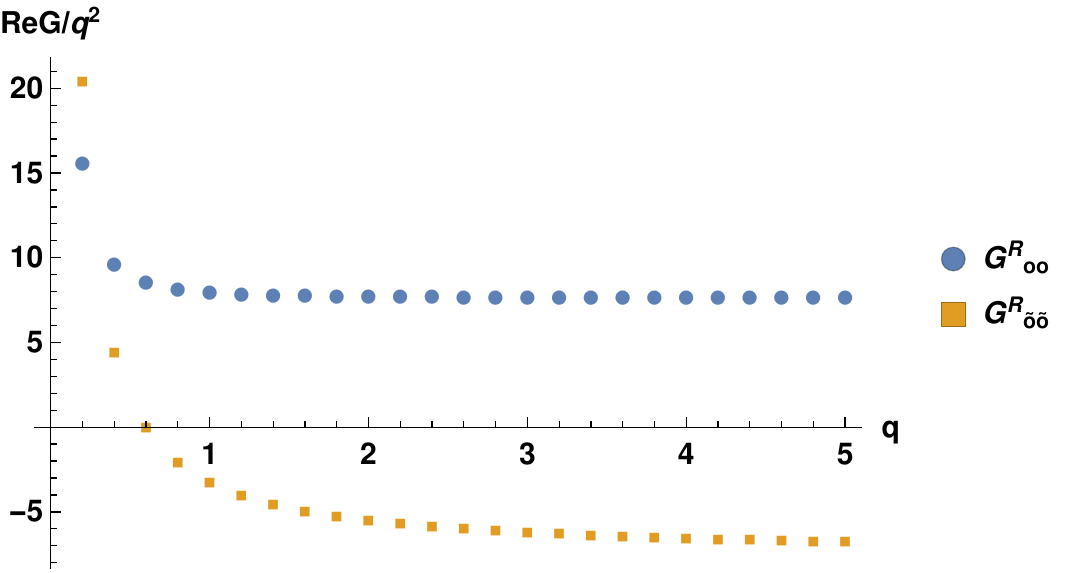}
\includegraphics[width=6cm]{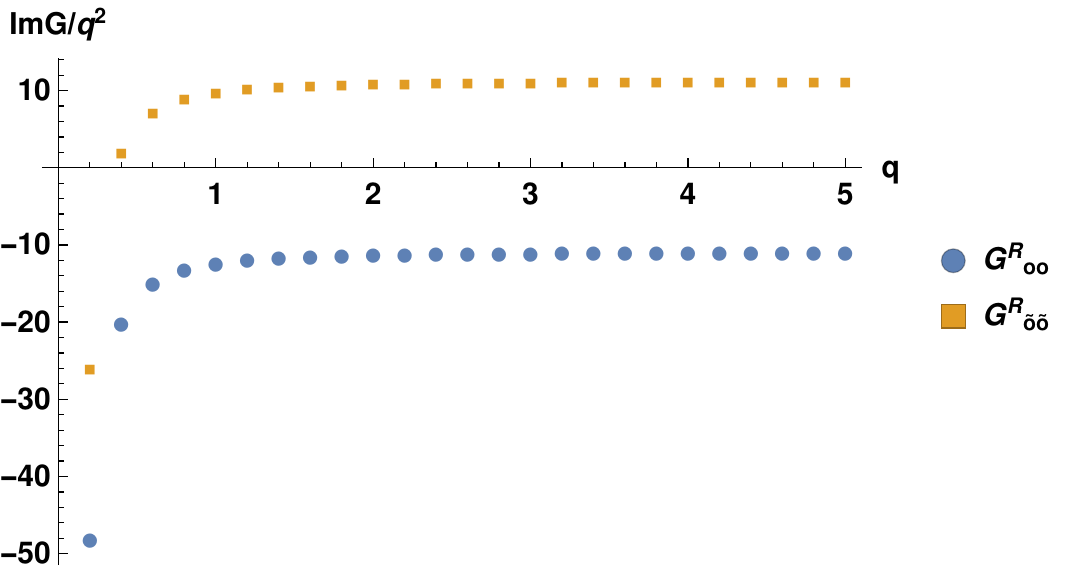}
\caption{\label{fig:4}The top panel show the real and imaginary parts of $G^R_{OO}$ and $G^R_{\tilde{O}\tilde{O}}$ in unit of $P$ at $q/\left(\pi T\right)^4=1$ as a function of $\omega$. At small $\o$, we recover the degeneracy of $G^R_{OO}$ and $-G^R_{\tO\tO}$ found before. At large $\o$, the plot reveal the degeneracy of $G^R_{OO}$ and $G^R_{\tO\tO}$ instead as expected on general ground. The bottom panels show the $q$ dependence of the real and imaginary parts of $G^R_{OO}$ and $G^R_{\tO\tO}$ in unit of $P$ at $\o/\(\pi T\)=1$. At large $q$, numerical results suggest a $q^2$ dependence. Note that this regime is close to the degeneracy regime of $G^R_{OO}$ and $-G^R_{\tO\tO}$.}
\end{figure}

Moving on from the regime $\o\ll T$, we study wider range of $\o$. We expect when $\o\gg q$, the instanton effect becomes negligible, thus $G^R_{OO}$ and $G^R_{\tO\tO}$ again becomes degenerate. We show the $\o$ and $q$ dependence of $G^R_{OO}$ and $G^R_{\tO\tO}$ in wider range of $\o$. The degeneracy of $G^R_{OO}$ and $-G^R_{\tO\tO}$ in small $\o$ regime and the degeneracy $G^R_{OO}$ and $G^R_{\tO\tO}$ in large $\o$ regime are visible in the Figure \ref{fig:4}. The figure also suggests the saturation of $q^2$ dependence for both real and imaginary parts of $G^R_{OO}$ and $G^R_{\tO\tO}$.

\section{Discussion}
 
The regime with $\o\ll T$ and $q\sim O(T^4)$ deserves special attention. Numerical results suggest the scaling \eqref{scaling}, which after reinstating dimension gives
\begin{align}\label{scaling_dim}
ImG^R_{OO}(\o)\simeq -ImG^R_{\tO\tO}(\o)= -\frac{q^2\G}{\o},
\end{align}
where $\G\sim O\(\frac{1}{T^3}\)$. From Fig \ref{fig:2} we can see $\G$ is positive. Note that the imaginary part of retarded correlator is related to the spectral density $\c_O$ and $\c_{\tO}$ as
\begin{align}
\c_O=-2ImG^R_{OO},\quad \c_{\tO}=-2ImG^R_{\tO\tO}.
\end{align}
Spectral density should be positive for positive frequency $\o$ \cite{Bellac:2011kqa}. Combining with \eqref{scaling_dim}, we find indeed positive spectral density for $O$, but negative spectral density for $\tO$. Away from the regime, we do recover positive spectral density for both $O$ and $\tO$ from Fig \ref{fig:4}. We suggest the violation of the positive condition for $\tO$ in the regime is related to the existence of unstable fluctuation of $\tO$. To see this, we consider the following correlator
\begin{align}\label{to_fluc}
\int d^3x\lag\(\tO(t,x)-\tO(0)\)^2\rag.
\end{align}
Using Kubo-Martin-Schwinger (KMS) relation, we can express \eqref{to_fluc} in terms of imaginary part of retarded correlator
\begin{align}\label{growth}
\int d^3x\lag\(\tO(t,x)-\tO(0)\)^2\rag&=\int\frac{d\o}{2\pi}\(2-e^{-i\o t}-e^{i\o t}\)G_{\tO\tO}^>(\o) \no
&=\int\frac{d\o}{2\pi}\(2-e^{-i\o t}-e^{i\o t}\)\frac{2}{1-e^{-\b\o}}ImG_{\tO\tO}^R(\o) \no
&\simeq \int\frac{d\o}{2\pi}\(2-e^{-i\o t}-e^{i\o t}\)\frac{2q^2\G T}{\o^2}\sim q^2\G T t.
\end{align}
In the last step, we approximate the integrand using \eqref{scaling_dim}. This is justified as the dominant contribution at large $t$ comes from small $\o$. This gives the long time behavior of the fluctuation. The linear dependence in $t$ shows it is a random walk growth, which is unstable. 
Note that the scaling relation holds in the regime $\o\ll T$, which sets the time scale of the unstable mode. In other words, on a time scale $t\sim \frac{1}{T}$, fluctuation effectively destroys the D-instanton background. Therefore, we conclude that the lifetime of instanton is set by the temperature. The conclusion is in line with the classic field theory results that large instanton is suppressed at high temperature \cite{Pisarski:1980md,Shuryak:1994ay}. As a final remark, we stress that classic results were obtained based on thermodynamics consideration, our conclusion were obtained from analysis of real time fluctuations.

\section{Acknowledgments}

SL is grateful to Yan Liu and Nan Su for valuable discussions. SL is in part supported by One Thousand Talent Program for Young Scholars and NSFC under Grant No 11675274 and NSFC Key Project under Grant No 11735007. SWL is in part supported by Fudan University and One Thousand Talent Program for Young Scholars.

\appendix

\section{Holographic renormalization}

The purpose of this appendix is to express correlation functions
in terms of elements of the response matrix. The derivation closely follows \cite{Policastro:2002tn,Kaminski:2009dh}.
We start from the 5D
Euclidean action (\ref{eq:9}) with Gibbon-Hawking term $S_{GH}$\footnote{To simplify the calculation, we have set the 5d cosmological $\Lambda=-6$ and assume all the fields depend on $\t$ and $r$ only, which is case of our interest.},
\begin{align}
S_{E}= & S_{5D}+S_{GH},\nonumber \\
= & \frac{1}{\k_5^2}\int d^{5}x\sqrt{g_{(5)}}\left[\mathcal{R}^{(5)}-\frac{1}{2}\left(\partial\Phi\right)^{2}+\frac{1}{2}e^{2\Phi}\left(\partial\chi\right)^{2}+12\right]+\frac{2}{\k^2}\int d^{4}x\sqrt{h}K.
\end{align}
Following \cite{Liu:1998bu}, we add an additional counter
term $S_{CT}$ to cancel the volume divergence to the bulk theory
which is,
\begin{equation}
S_{CT}=-\frac{3}{\k_5^2}\int d^{4}x\sqrt{h}.
\end{equation}
Note that this is the counter term for gravity sector. Other counetr terms for dilaton and axion are also needed, see \cite{Gutperle:2002km} and references therein for early constructions. We will need additional counter terms in the discussion below.
We assume the boundary is taken at $r=\Sigma$ which would be sent
to infinity at the end of the calculation. Using the equations of
motion in (\ref{eq:11}) and keeping the contribution from the boundary
$r=\Sigma$, and combining the resultant boundary terms with $S_{GH}$
and $S_{CT}$, we can obtain,
\begin{align}
S_{E}= &\frac{1}{2\k_5^2}\int d^4x \bigg[\frac{3r^{4}h_{00}^{2}}{4f^{3/2}}\left(1-f^{1/2}\right)+\frac{9r^{4}Hh_{00}}{2f^{1/2}}\left(f^{1/2}-1\right)-\frac{9r^{4}H^{2}}{4}\left(f^{1/2}-1\right)\nonumber \\
 & -\frac{3r^{5}f^{\prime}Hh_{00}}{8f}+\frac{3r^{5}f^{\prime}H^{2}}{8}-\frac{1}{4}r^{5}\left(h_{00}+3fH\right)\delta\Phi\Phi^{\prime}+\frac{1}{4}e^{\Phi}r^{5}\left(h_{00}+3fH\right)\delta\chi\Phi^{\prime}\nonumber \\
 & +e^{\Phi}r^{5}f\delta\Phi\delta\chi\Phi^{\prime}+\frac{3}{4}r^{5}\left(Hh_{00}\right)'+\frac{3}{2}r^{5}fH^{\prime}H-\frac{1}{2}r^{5}f\delta\Phi\(\delta\Phi\)'+\frac{1}{2}e^{2\Phi}r^{5}\delta\chi\(\delta\chi\)'\bigg].
\end{align}
Plugging the asymptotic expansions (\ref{eq:16}) into the above formula,
we obtain the on-shell action,
\begin{align}
S_{E}= &\frac{1}{2\k_5^2}\int d^4x \bigg[ \left(-\frac{3a_{1}b_{0}}{2}-c_{0}c_{1}+f_{0}f_{1}\right)r^{2}+\frac{3}{8}a_{0}^{2}-\frac{15}{4}a_{0}b_{0}-3a_{2}b_{0}+\frac{3}{8}b_{0}^{2}-3a_{0}b_{2}\nonumber \\
 &-6b_{0}b_{2}-c_{1}^{2} -2c_{0}c_{1}+\frac{1}{2}c_{0}c_{h}+f_{1}^{2}+2f_{0}f_{2}-\frac{1}{2}f_{0}f_{h}+q\left(-a_{0}c_{0}-3b_{0}c_{0}+a_{0}f_{0}\right. \nonumber \\
 & \left.+3b_{0}f_{0}-4c_{0}f_{0}\right)-16\left(c_{0}c_{h}-f_{0}f_{h}\right)\ln r+...\bigg]
\end{align}
Note that all the coefficients are functions of $t$: $a_0=a_0(t)$ etc. To evaluate the correlators in momentum space, we express $S_E$ in terms of Fourier components of the coefficient $a_0(k)$. In doing this, we find that the coefficients of the superficially $r^{2}$ divergent terms are of the type $\oE^2b_0^2$, $\oE^2c_0^2$ and $\oE^2f_0^2$. Similarly, the $\ln r$ divergent terms are of the type $\oE^4c_0^2$ and $\oE^4f_0^2$. Therefore all divergent terms are contact terms thus should be discarded \cite{Kovtun:2006pf}. The remaining action is given by
%
\begin{align}\label{eq:26}
S_{E}&=  \frac{1}{2\k_5^2(2\pi)}\int d\oE \bigg[ -3b_{0}(-\oE)a_{2}(\oE)
-6b_{0}(-\oE)b_{2}(\oE)-2c_{0}(-\oE)c_{2}(\oE) \no
&+2f_{0}(-\oE)f_{2}(\oE)+\frac{3}{8}\big[a_{0}(-\oE)a_0(\oE)-10a_{0}(-\oE)b_{0}(\oE)+3b_{0}(-\oE)b_0(\oE)\big]\nonumber \\
&+q\big[-a_{0}(-\oE)c_{0}(\oE)-3b_{0}(-\oE)c_{0}(\oE)+a_{0}(-\oE)f_{0}(\oE)+3b_{0}(-\oE)f_{0}(\oE) \no
&-4c_{0}(-\oE)f_{0}(\oE)\big]\bigg].
\end{align}
There are three types of terms in \eqref{eq:26}. The first type is product of sources and VEVs like $b_0(-\oE)a_2(\oE)$. The second type is product of sources like $a_0(-\oE)a_0(\oE)$, which indicates the presence of constant terms in the resulting correlators.
The third type is product of sources and $q$, such as $qa_0(-\oE)c_0(\oE)$.
Now we use the response matrix to express VEVs in terms of sources. We can then proceed to calculate the correlators by differentiation the action with respect to sources.
\begin{align}\label{differentiation}
G_{00,00}^E&=\int d\tau d^{3}xe^{i\omega_{E}\tau}\left\langle T_{E}^{00}\left(\tau,x\right)T_{E}^{00}\left(0\right)\right\rangle = 4\frac{(2\pi)\d^2S_E}{\d a_0(-\oE)\d a_0(\oE)},\nonumber\\
G_{00,ii}^E&=\int d\tau d^{3}xe^{i\omega_{E}\tau}\left\langle T_{E}^{00}\left(\tau,x\right)T_{E}^{ii}\left(0\right)\right\rangle = 4\frac{(2\pi)\d^2S_E}{\d a_0(-\oE)\d b_0(\oE)},\nonumber\\
G_{ii,jj}^E&=\int d\tau d^{3}xe^{i\omega_{E}\tau}\left\langle T_{E}^{ii}\left(\tau,x\right)T_{E}^{jj}\left(0\right)\right\rangle = 4\frac{(2\pi)\d^2S_E}{\d b_0(-\oE)\d b_0(\oE)},\nonumber\\
G_{00,O}^E&=\int d\tau d^{3}xe^{i\omega_{E}\tau}\left\langle T_{E}^{00}\left(\tau,x\right)O_{E}\left(0\right)\right\rangle =  2\frac{(2\pi)\d^2S_E}{\d a_0(-\oE)\d f_0(\oE)},\nonumber \\
G_{00,\tO}^E&=\int d\tau d^{3}xe^{i\omega_{E}\tau}\left\langle T_{E}^{00}\left(\tau,x\right)\tilde{O}_{E}\left(0\right)\right\rangle = 2\frac{(2\pi)\d^2S_E}{\d a_0(-\oE)\d c_0(\oE)},\nonumber \\
G_{ii,O}^E&=\int d\tau d^{3}xe^{i\omega_{E}\tau}\left\langle T_{E}^{ii}\left(\tau,x\right)O_{E}\left(0\right)\right\rangle = 2\frac{(2\pi)\d^2S_E}{\d b_0(-\oE)\d f_0(\oE)},\nonumber \\
G_{ii,\tO}^E&=\int d\tau d^{3}xe^{i\omega_{E}\tau}\left\langle T_{E}^{ii}\left(\tau,x\right)\tilde{O}_{E}\left(0\right)\right\rangle = 2\frac{(2\pi)\d^2S_E}{\d b_0(-\oE)\d c_0(\oE)},\nonumber \\
G_{OO}^E&=\int d\tau d^{3}xe^{i\omega_{E}\tau}\left\langle O_{E}\left(\tau,x\right)O_{E}\left(0\right)\right\rangle = \frac{(2\pi)\d^2S_E}{\d f_0(-\oE)\d f_0(\oE)},\nonumber \\
G_{\tO\tO}^E&=\int d\tau d^{3}xe^{i\omega_{E}\tau}\left\langle \tilde{O}_{E}\left(\tau,x\right)\tilde{O}_{E}\left(0\right)\right\rangle = \frac{(2\pi)\delta^{2}S}{\delta c_{0}(-\oE)\delta c_0(\oE)},\nonumber \\
G_{O\tO}^E&=\int d\tau d^{3}xe^{i\omega_{E}\tau}\left\langle O_{E}\left(\tau,x\right)\tilde{O}_{E}\left(0\right)\right\rangle = \frac{(2\pi)\delta^{2}S}{\delta f_{0}(-\oE)\delta c_{0}(\oE)}.
\end{align}
We have suppressed the overall pre-factor $\frac{1}{2\k_5^2}$ for notational simplicity. Note that the pre-factor has mass dimension four. We restore dimension by multiplying $(\pi T)^4$ and convert the pre-factor to field theory quantity $\frac{(\pi T)^4}{2\k_5^2}=\frac{\pi^2N^2T^4}{8}=P$, which is the same as pressure of plasma.
Recall that we have confirmed the response matrix is even in $\oE$. This nice property allows to treat the sources as ordinary numbers. This is essentially the procedure adopted in \cite{Policastro:2002tn}.
When the response matrix is not even in $\oE$, a more careful treatment is needed \cite{Kaminski:2009dh}. Plugging \eqref{eq:21} into \eqref{differentiation} we obtain \eqref{eq:28}.

\bibliographystyle{unsrt}
\bibliography{Dinstanton}
\end{document}